\documentclass[a4paper,spanish,english,manuscript]{revtex4}
\usepackage[T1]{fontenc}
\usepackage[latin1]{inputenc}
\usepackage{textcomp}
\usepackage{amsmath}
\usepackage{amssymb}
\usepackage{esint}

\makeatletter

\@ifundefined{textcolor}{}
{%
 \definecolor{BLACK}{gray}{0}
 \definecolor{WHITE}{gray}{1}
 \definecolor{RED}{rgb}{1,0,0}
 \definecolor{GREEN}{rgb}{0,1,0}
 \definecolor{BLUE}{rgb}{0,0,1}
 \definecolor{CYAN}{cmyk}{1,0,0,0}
 \definecolor{MAGENTA}{cmyk}{0,1,0,0}
 \definecolor{YELLOW}{cmyk}{0,0,1,0}
}

\usepackage{babel}
\addto\shorthandsspanish{\spanishdeactivate{~<>}}

\usepackage{babel}

\makeatother

\usepackage{babel}
\addto\shorthandsspanish{\spanishdeactivate{~<>}}

\begin{document}

\title{Entropy Production in Simple Special Relativistic Fluids}

\author{D. Brun-Battistini$^{1}$, A. Sandoval-Villalbazo$^{1}$, A. L. Garcia-Perciante$^{2}$}

\address{$^{1}$Depto. de Fisica y Matematicas, Universidad Iberoamericana,
Prolongacion Paseo de la Reforma 880, Mexico D. F. 01219, Mexico.}

\address{$^{2}$Depto. de Matematicas Aplicadas y Sistemas, Universidad Autonoma
Metropolitana-Cuajimalpa, Prol. Vasco de Quiroga 4871, Mexico D.F
05348, Mexico.\\
 }
\begin{abstract}
It is well known that, in the absence of external forces, simple non-relativistic
fluids involve entropy production only through heat conduction and
shear viscosity \cite{key-1}. In this work, it is shown that a number
density gradient contributes to the local entropy production of a
simple relativistic fluid using special relativistic kinetic theory.
Also, the presence of an external field may cause strictly relativistic
contributions to the entropy production, a fact not widely recognized.
The implications of these effects are thoroughly discussed.
\end{abstract}
\maketitle

\section{Introduction}

Force-flux relations in irreversible thermodynamics are phenomenologically
proposed as to comply with the second law of thermodynamics. In the
usual procedure, one uses the local equilibrium assumption together
with the functional dependence $s\left(\vec{r},t\right)=s\left(n\left(\vec{r},t\right),\varepsilon\left(\vec{r},t\right)\right)$
in order to write a balance equation for the entropy density $s$
\cite{key-1}. This functional dependence arises from the fact that
the most natural choice for the independent scalar state variables
is the particle number density $n$ and the internal energy density
$\varepsilon$. Through them one can establish the space and time
dependence of the rest of the thermodynamical variables. In the non-relativistic
case the entropy balance equation has the general form 
\begin{equation}
n\dot{s}+\nabla\cdot\vec{J}_{s}=\sigma\label{eq:1-1}
\end{equation}
where $\vec{J}_{s}$ is the entropy flux (related with the heat flux).
The entropy production $\sigma$ is then rewritten, after introducing
the transport equations for $n$ and $\varepsilon$, in the following
form
\begin{equation}
\sigma=\sum\vec{J}_{i}\cdot\vec{F}_{i}\label{eq:2-1}
\end{equation}
where $\vec{J}_{i}$ are the still unknown fluxes and $\vec{F}_{i}$
the corresponding thermodynamic forces. In this case the thermodynamic
forces, as gradients of the state variables, arise naturally as well
as the flux-force relations proposed in order to satisfy the extension
of the second law of thermodynamics for irreversible processes, $\sigma\geq0$.
These are linear relations of the Fourier type, relating the heat
flux with the temperature gradient in a simple ideal fluid. 

However for high temperature gases, in which relativistic effects
in the dynamics of individual molecules are relevant, the transport
equations feature new terms \cite{Eckart}. The introduction of such
expressions in the entropy balance yields an equation in which the
choice of thermodynamic forces is not as direct as in the non-relativistic
case. Based on such equation, Ref. \cite{Eckart} proposes the
hydrodynamic acceleration as a thermodynamic force corresponding to
the heat flux. However, it has been shown that such choice leads to
the so-called generic instabilities which in part led to the conclusion
that second order type theories were required in order to properly
describe the behavior of special relativistic gases \cite{Hiscock-Lindblom,Onthe nature of the so called generic instabilities in dissipative relativistic hydrodynamics}. 

On the other hand, kinetic theory provides a framework for establishing
from first principles the relativistic transport equations for the
state variables, together with expressions for the corresponding fluxes
and production terms. Using these equations and completing the system
with the constitutive equations obtained by solving Boltzmann's equation
to first order in the gradients, one obtains a set of hydrodynamic
equations that yields no generic instabilities, and is thus consistent
with the Onsager regression of fluctuations hypothesis \cite{Physica A 2009}\cite{ONSG}.
When this procedure is carried out, one encounters that the force-flux
relations feature strictly relativistic effects that are not present
in the non-relativistic simple gas and, in some cases, have not been
obtained phenomenologically. Such is the case of the coupling of the
heat flux with the chemical potential, density or pressure gradient
(see Refs. \cite{israel63},\cite{Garcia-Perciante Mendez Heat conduction ... revisited}
and \cite{ck} respectively) which phenomenologically is usually written
in terms of the hydrodynamic acceleration. In Ref. \cite{benedicks}
it has been shown, by applying the standard techniques of relativistic
kinetic theory to the case of a single component charged gas in the
precense of an electrostatic field, that the heat flux has an additional
driving force given by the gradient of the electrostatic potential.
The purpose of this paper is to establish the entropy production for
such a case and establish a link to the fluxes-forces formalism. The
relation of the production term with the electrostatic field is then
identified with Benedicks (thermoelectric) effect \cite{benoriginal}.

To accomplish this task we have divided this paper as follows. Section
II is devoted to establishing the balance equation for the entropy
density and the corresponding expression for the entropy production,
consistent with the second law. The explicit calculation of the entropy
production is shown in Section III. The discussion of the results
and final conclusions, as well as possible generalizations, are included
in Section IV.

\section{The relativistic entropy balance equation}

The starting point for establishing an entropy balance equation for
a relativistic gas is, as in the non-relativistic case, Boltzmann's
equation. For a high temperature, dilute, non-degenerate gas such
equation is given by \cite{degroor,ck} 
\begin{equation}
v^{\alpha}\frac{\partial f}{\partial x^{\alpha}}+\dot{v}^{\alpha}\frac{\partial f}{\partial v^{\alpha}}=J\left(ff'\right)\label{eq:1}
\end{equation}
which corresponds to an evolution equation in phase space for the
single-particle distribution function. Here $v^{\alpha}$ is the molecular
four-velocity, $x^{\alpha}$ the space-time position four-vector in
a Minkowski metric $\eta^{\alpha\beta}$ with a $+++-$ signature
and $J\left(ff'\right)$ is the collision term \cite{degroor}. The
distribution function $f$ has the same interpretation as in the non-relativistic
case, that is a function such that 
\begin{equation}
f\left(x^{\nu},v^{\nu}\right)d^{3}xd^{3}v\label{eq:1.5}
\end{equation}
gives the number of points in a cell in the six dimensional phase
space with volume $d^{3}xd^{3}v$. The second term on the left hand
side of Eq. (\ref{eq:1}) contains the four-acceleration \emph{of
a single particle} which, for the case of molecules carrying charge
$q$ and having rest mass $m$ can be written as \cite{key-4}
\begin{equation}
\dot{v}^{\mu}=\frac{q}{m}v_{\nu}F^{\mu\nu}\label{eq:2}
\end{equation}
Here $F^{\mu\nu}$ is the electromagnetic field tensor which for an
electrostatic field is given by 
\begin{equation}
F^{\mu\nu}=\left(\begin{array}{cccc}
0 & 0 & 0 & -\frac{\phi^{,1}}{c}\\
0 & 0 & 0 & -\frac{\phi^{,2}}{c}\\
0 & 0 & 0 & -\frac{\phi^{,3}}{c}\\
\frac{\phi^{,1}}{c} & \frac{\phi^{,2}}{c} & \frac{\phi^{,3}}{c} & 0
\end{array}\right)\label{eq:3}
\end{equation}
where $\phi$ is the electrostatic potential and $c$ the speed of
light. 

In order to obtain the entropy balance equation, we follow the standard
procedure for obtaining transport equations for the state variables
\cite{ck}. The Boltzmann equation is multiplied by $k_{b}\ln f$,
where $k_{b}$ is Boltzmann's constant, and integrated over velocity
space using the invariant volume element $d^{*}v=\frac{d^{3}v}{v^{4}}$
. In this case the acceleration term vanishes and one is led to (see
Appendix A)

\begin{equation}
-k_{b}\frac{\partial}{\partial x^{\mu}}\int v^{\mu}f\left(\ln f\right)d^{*}v=-k_{b}\int J\left(ff'\right)\ln fd^{*}v\label{eq:3.1}
\end{equation}
which can be written in a more familiar way as 
\begin{equation}
\frac{\partial S^{\nu}}{\partial x^{\nu}}=\sigma\label{eq:3.2}
\end{equation}
where $S^{\nu}$ is the entropy four-flux and $\sigma$ is the entropy
production. These quantities are given by
\begin{equation}
S^{\mu}=-k_{b}\int v^{\mu}f\left(\ln f\right)d^{*}v\label{eq:3.3}
\end{equation}
and
\begin{equation}
\sigma=-k_{b}\int J\left(ff'\right)\ln fd^{*}v\label{eq:3.4}
\end{equation}
respectively. From Eq. (\ref{eq:3.2}) one can clearly recognize that
the extension of the second law of thermodynamics to irreversible
processes is given by
\begin{equation}
\sigma\geq0\label{eq:3.5}
\end{equation}

The next step in order to obtain the entropy production is to introduce
Chapman-Enskog's hypothesis for the distribution function, that is
\begin{equation}
f=f^{\left(0\right)}\left(1+\varphi+...\right)\label{eq:3.6}
\end{equation}
where $f^{\left(0\right)}\varphi$ is a first order in the gradients
correction to the local equilibrium distribution function $f^{\left(0\right)}$,
which for a relativistic gas is given by \cite{JUTTNER} 
\begin{equation}
f^{\left(0\right)}=\frac{n}{4\pi c^{3}z\mathcal{K}_{2}\left(\frac{1}{z}\right)}\exp\left(\frac{\mathcal{U}^{\beta}v_{\beta}}{zc^{2}}\right)\label{eq:4}
\end{equation}
Although the general expression involves an infinite sum, retaining
only the first correction corresponds to Navier-Stokes regime which
is the main interest of this work. Substituting Eq. (\ref{eq:3.6})
in the expression for the entropy production given by Eq. (\ref{eq:3.4})
leads to
\begin{equation}
\sigma=-k_{b}\int J\left(ff'\right)\ln\left(1+\varphi\right)d^{*}v\label{eq:3.7}
\end{equation}
Notice that the term depending on $f^{\left(0\right)}$ vanishes upon
integration, being the distribution function a collisional invariant
within the Boltzmann equation's hypothesis. Also, since the correction
is assumed to be small we have $\ln\left(1+\varphi\right)\sim\varphi$
and thus one can write
\begin{equation}
\sigma=-k_{b}\int J\left(ff'\right)\varphi d^{*}v\label{eq:3.8}
\end{equation}
For the sake of simplicity, we now introduce the relaxation time approximation
for the collision term, proposed by Marle \cite{ck}, and write
\begin{equation}
J\left(ff'\right)=-\frac{f-f^{\left(0\right)}}{\tau}\label{eq:3.9}
\end{equation}
Introducing Eqs. (\ref{eq:3.6}) and (\ref{eq:3.9}) in Eq. (\ref{eq:3.8})
one obtains for the entropy production
\begin{equation}
\sigma=\frac{k_{b}}{\tau}\int f^{\left(0\right)}\varphi^{2}d^{*}v\label{eq:7}
\end{equation}
which is consistent with the extension of the second law (see Eq.
(\ref{eq:3.5})). It is worthwhile noticing that the previous expression
only holds in the case of a relaxation time approximation. The general
expression for the entropy production in the Chapman-Enskog expansion
is given by (\ref{eq:3.8}).

\section{Calculation of the entropy production}

The entropy production can be explicitly calculated from Eq. (\ref{eq:3.8})
in the Chapman-Enskog approximation by noticing that the collision
kernel can be expressed as \cite{key-1}
\begin{equation}
J\left(ff'\right)=\int\left(f'_{1}f'-f_{1}f\right)\mathcal{F}\sigma d\Omega dv_{1}^{*}=J^{\left(0\right)}+J^{\left(1\right)}+J^{\left(2\right)}\label{eq:85}
\end{equation}
where the ordering corresponds to ascending powers of the Knudsen
parameter, roughly the order in the gradients of the thermodynamic
variables. The terms in Eq. (\ref{eq:85}) are given by 
\[
J^{\left(0\right)}=0
\]
\[
J^{\left(1\right)}=\int f'{}_{1}^{\left(0\right)}f'{}^{\left(0\right)}\left(\varphi'+\varphi'_{1}-\varphi_{1}-\varphi\right)\mathcal{F}\sigma d\Omega dv_{1}^{*}
\]
\[
J^{\left(2\right)}=\int f'{}_{1}^{\left(0\right)}f'{}^{\left(0\right)}\left(\varphi'\varphi'_{1}-\varphi_{1}\varphi\right)\mathcal{F}\sigma d\Omega dv_{1}^{*}
\]
where the order zero approximation vanishes since $f'{}_{1}^{\left(0\right)}f'{}^{\left(0\right)}=f_{1}^{\left(0\right)}f^{\left(0\right)}$.
Thus, the entropy production corresponding to the first order in the
gradients Chapman-Enskog approximation is given by
\begin{equation}
\sigma^{\left(1\right)}=-k_{b}\int f_{1}'{}^{\left(0\right)}f'^{\left(0\right)}\left(\varphi'+\varphi'_{1}-\varphi_{1}-\varphi\right)\varphi\mathcal{F}\sigma d\Omega dv_{1}^{*}d^{*}K\label{eq:55}
\end{equation}
Expressing the collision kernel $J^{\left(1\right)}$ in terms of
the left hand side of Boltzmann's equation in the corresponding order,
one obtains

\begin{equation}
\sigma^{\left(1\right)}\simeq-k_{b}\int\left[v^{\alpha}f_{,\alpha}^{\left(0\right)}+\frac{q}{m}v_{\alpha}F^{\mu\alpha}\frac{\partial f^{\left(0\right)}}{\partial v^{\mu}}\right]\varphi d^{*}K\label{eq:9}
\end{equation}
The term in brackets has already been calculated elsewhere (see equations
(7) and (27) in Ref. \cite{benedicks}) and can be written as 
\begin{align}
v^{\alpha}f_{,\alpha}^{\left(0\right)}+\frac{q}{m}v_{\alpha}F^{\mu\alpha}\frac{\partial f^{\left(0\right)}}{\partial v^{\mu}} & =f^{\left(0\right)}\gamma_{\left(k\right)}h_{\nu}^{\mu}k^{\nu}\left\{ \frac{1}{z}\left(\frac{\gamma_{\left(k\right)}}{\mathcal{G}\left(\frac{1}{z}\right)}-1\right)\frac{q}{mc^{2}}\phi_{,\mu}+\left(1-\frac{\gamma_{\left(k\right)}}{\mathcal{G}\left(\frac{1}{z}\right)}\right)\frac{n_{,\mu}}{n}\right.\nonumber \\
 & \left.+\left(1-\frac{\gamma_{\left(k\right)}}{z}-\frac{\gamma_{\left(k\right)}}{\mathcal{G}\left(\frac{1}{z}\right)}-\frac{\mathcal{G}\left(\frac{1}{z}\right)}{z}\right)\frac{T_{,\mu}}{T}\right\} \label{eq:27}
\end{align}
where $k^{\ell}$ is the peculiar velocity, defined as the single
molecule velocity measured by an observer comoving with the fluid.
The speed dependence is included in $\gamma_{\left(k\right)}$, given
by
\begin{equation}
\gamma_{\left(k\right)}=-\frac{\mathcal{U}^{\mu}v_{\mu}}{c^{2}}\label{eq:10}
\end{equation}
which is the usual Lorentz factor
\begin{equation}
\gamma_{\left(k\right)}=\left(1-\frac{k^{2}}{c^{2}}\right)^{-1/2}\label{eq:11}
\end{equation}
with $k$ being the magnitude of $k^{\ell}$. Also, the spatial projector
$h^{\mu\nu}$ in Eq. (\ref{eq:27}) is given by
\begin{equation}
h^{\mu\nu}=\eta^{\mu\nu}+\frac{\mathcal{U}^{\mu}\mathcal{U}^{\nu}}{c^{2}}\label{eq:12}
\end{equation}
Substituting Eq. (\ref{eq:27}) in Eq. (\ref{eq:9}) and recalling
that the spatial components of the relativistic heat flux are given
by \cite{nosJNET-1} 
\begin{equation}
J_{\left[Q\right]}^{\mu}=mc^{2}h_{\nu}^{\mu}\int k^{\nu}f^{\left(1\right)}\gamma_{\left(k\right)}^{2}d^{*}K\label{eq:13}
\end{equation}
one obtains (see Appendix B)
\begin{equation}
\sigma^{\left(1\right)}=-J_{\left[Q\right]}^{\ell}\cdot\left\{ \left(1-\frac{z}{\mathcal{G}\left(\frac{1}{z}\right)}\right)\frac{T_{,\ell}}{T^{2}}-\frac{z}{\mathcal{G}\left(\frac{1}{z}\right)T}\frac{n_{,\ell}}{n}+\frac{z}{\mathcal{G}\left(\frac{1}{z}\right)}\frac{q}{k_{b}T^{2}}\phi_{,\ell}\right\} \label{eq:14}
\end{equation}
This is the final result of this work which shows that there are three
relativistic contributions to the entropy production in an inhomogeneous
charged gas in the presence of an electrostatic field, namely the
usual Fourier term, the relativistic contribution due to the particle
number density gradient and the Benedicks-like term.

\section{Final remarks}

The phenomenological approach to the establishment of a positive semidefinite
local entropy production in a simple relativistic fluid suggests a
constitutive equation that couples the heat flux with the hydrodynamic
acceleration \cite{Eckart}. Nevertheless, this coupling leads to
the so-called generic instabilities first identified by Hiscock and
Lindlblom back in 1985 \cite{Hiscock-Lindblom}. In this context,
a consititutive equation for the heat flux that solely includes spatial
first order gradients of the local thermodynamic variables was established
based on the grounds of relativistic kinetic theory. This constitutive
equation \emph{solves the generic instabilities problem} \cite{Physica A 2009}.
Following these ideas, the next logical step corresponds to the analysis
of the entropy production based on kinetic theory, so that it can
be compared with its phenomenological counterpart.

The main results of this work, basically contained in Eqs. (\ref{eq:7})
and (\ref{eq:14}), show that a positive semidefinite expression for
the entropy production is obtained using standard kinetic theory arguments.
This expression includes only first order gradients in the local variables
and reduces to the classical expression in the limit $z\rightarrow0$.
Equation (\ref{eq:14}) also shows that in the $z\simeq1$ regime,
a Benedicks-type effect arises as a purely relativistic source of
entropy \cite{benedicks}. 

Finally, it is interesting to notice a subtle step involved in the
present derivation. The single particle acceleration term present
in the left hand side of Eq. (\ref{eq:1}) is given in covariant form
in Eq. (\ref{eq:2}), which is consistent with special relativity.
However, if a gravitational field is considered and general relativistic
effects become important, Eq. (\ref{eq:2}) must be replaced with
an expression that takes into account curvature effects in the dynamics
of the individual particles (a geodesic for example). In this case
it is still unclear if the entropy production would show similar features
as those included in Eq. (\ref{eq:14}) {[}15-16{]}. This subject
will be addressed in the near future.

\appendix

\section*{Appendix A}

In order to obtain the entropy balance equation, Eq. (\ref{eq:3.1}),
one multiplies Boltzmann's equation (\ref{eq:1-1}) by $k_{b}\ln f$
and integrates over velocity space. That is
\begin{equation}
\frac{\partial}{\partial x^{\alpha}}k_{b}\int v^{\alpha}f\ln fd^{*}v+\frac{q}{m}k_{b}\int v_{\alpha}F^{\mu\alpha}\ln f\frac{\partial f}{\partial v^{\alpha}}d^{*}v=k_{b}\int J\left(ff'\right)\ln fd^{*}v\label{eq:s1}
\end{equation}
where in the first term on the left hand side use has been made of
the fact that the distribution function only depends on space-time
coordinates through the state variables. For the acceleration term,
one can use
\begin{equation}
\frac{\partial}{\partial v^{\alpha}}\left(v_{\beta}f\ln f\right)=v_{\beta}f\frac{\partial}{\partial v^{\alpha}}\left(\ln f\right)+v_{\beta}\ln f\frac{\partial f}{\partial v^{\alpha}}+\eta_{\beta\alpha}f\ln f\label{eq:s2}
\end{equation}
in order to write
\begin{equation}
\int v_{\beta}F^{\alpha\beta}\ln f\frac{\partial f}{\partial v^{\alpha}}d^{*}v=F^{\alpha\beta}\left[\int\frac{\partial}{\partial v^{\alpha}}\left(v_{\beta}f\ln f\right)d^{*}v-\int v_{\beta}f\frac{\partial}{\partial v^{\alpha}}\left(\ln f\right)d^{*}v-\int\eta_{\beta\alpha}f\ln fd^{*}v\right]\label{eq:s3}
\end{equation}
The first term in bracketts vanishes since the distribution function
tends to zero in the limits of integration. For the second term we
have, integrating by parts
\begin{equation}
\int v_{\beta}\frac{\partial f}{\partial v^{\alpha}}d^{*}v=-\eta_{\beta\alpha}n\label{eq:s4}
\end{equation}
 and thus
\begin{equation}
\int v_{\beta}F^{\alpha\beta}\ln f\frac{\partial f}{\partial v^{\alpha}}d^{*}v=F^{\alpha\beta}\eta_{\beta\alpha}\left(n-\int f\ln fd^{*}v\right)\label{eq:s5}
\end{equation}
which vanishes since $F^{\alpha\beta}$ is antisymmetric and thus
$F^{\alpha\beta}\eta_{\beta\alpha}=F_{\alpha}^{\alpha}=0$ is antisymmetric.
Notice also that this is only valid when $F^{\alpha\beta}$ does not
depend on the molecular velocity.

\section*{Appendix B}

In this appendix, the structure of the entropy production given by
Eq. (\ref{eq:14}) is established by substituting the left hand side
of the linearized Boltzmann's equation in the expression for $\sigma^{\left(1\right)}$
given by Eq. (\ref{eq:9}) which yields

\begin{align}
\sigma^{\left(1\right)} & \simeq k_{b}\int f^{\left(0\right)}\gamma_{\left(k\right)}h_{i}^{\ell}k^{i}\left\{ \frac{1}{z}\left(\frac{\gamma_{\left(k\right)}}{\mathcal{G}\left(\frac{1}{z}\right)}-1\right)\frac{q}{mc^{2}}\phi_{,\ell}+\left(1-\frac{\gamma_{\left(k\right)}}{\mathcal{G}\left(\frac{1}{z}\right)}\right)\frac{n_{,\ell}}{n}\right.\nonumber \\
 & =\left.\left(1-\frac{\gamma_{\left(k\right)}}{z}-\frac{\gamma_{\left(k\right)}}{\mathcal{G}\left(\frac{1}{z}\right)}-\frac{\mathcal{G}\left(\frac{1}{z}\right)}{z}\right)\frac{T_{,\ell}}{T}\right\} \varphi d^{*}K\label{eq:s6}
\end{align}
or, separating terms by the $\gamma_{\left(k\right)}$ dependence
\begin{align}
\sigma^{\left(1\right)} & \simeq k_{b}\left\{ \left[\frac{1}{\mathcal{G}\left(\frac{1}{z}\right)}\frac{q}{k_{b}T}\phi_{,\ell}-\frac{1}{\mathcal{G}\left(\frac{1}{z}\right)}\frac{n_{,\ell}}{n}-\left(\frac{1}{z}+\frac{1}{\mathcal{G}\left(\frac{1}{z}\right)}\right)\frac{T_{,\ell}}{T}\right]\int f^{\left(0\right)}k^{\ell}\gamma_{\left(k\right)}^{2}\varphi d^{*}K\right.\nonumber \\
 & +\left.\left[-\frac{q}{k_{b}T}\phi_{,\ell}+\frac{n_{,\ell}}{n}+\left(1-\frac{\mathcal{G}\left(\frac{1}{z}\right)}{z}\right)\frac{T_{,\ell}}{T}\right]\int f^{\left(0\right)}\gamma_{\left(k\right)}k^{\ell}\varphi d^{*}K\right\} \label{eq:s7}
\end{align}
Since
\begin{equation}
\int f^{\left(0\right)}k^{\ell}\gamma_{\left(k\right)}^{2}\varphi d^{*}K=\frac{J_{Q}^{\ell}}{mc^{2}}\label{eq:s8}
\end{equation}
and
\begin{equation}
\int f^{\left(0\right)}\gamma_{\left(k\right)}k^{\ell}\varphi d^{*}K=0\label{eq:s9}
\end{equation}
one obtains 
\begin{equation}
\sigma^{\left(1\right)}\simeq z\left[\frac{z}{\mathcal{G}\left(\frac{1}{z}\right)}\frac{q}{k_{b}T}\phi_{,\ell}-\frac{z}{\mathcal{G}\left(\frac{1}{z}\right)}\frac{n_{,\ell}}{n}-\left(1+\frac{z}{\mathcal{G}\left(\frac{1}{z}\right)}\right)\frac{T_{,\ell}}{T}\right]\frac{J_{Q}^{\ell}}{T}\label{eq:s10}
\end{equation}

\textsf{\textbf{\textit{\Large Acknowledgements}}}{\Large \par}

The authors acknowledge support from CONACyT through grant number
CB2011/167563.


\begin{thebibliography}{10}
\bibitem[1]{key-1}\foreignlanguage{spanish}{S.R. Groot de, P, Mazur;
\emph{Non-equilibrium Thermodynamics}, Dover Publications, Mineola,
N.Y., (1984).}

\selectlanguage{spanish}%
\bibitem[2]{Eckart} C. Eckart; \emph{The Thermodynamics of Irreversible
Processes. III}, J. Phys. Rev. \textbf{58}, 919-924 (1940).

\bibitem[3]{Hiscock-Lindblom} W. A. Hiscock, L. Lindblom; \emph{Generic
instabilities in first-order dissipative relativistic fluid theories},
Phys. Rev. D. \textbf{31} 4, 725 (1985).

\bibitem[4]{Onthe nature of the so called generic instabilities in dissipative relativistic hydrodynamics}
A.L. Garcia-Perciante, L.S. Garcia-Colin, Alfredo Sandoval-VIllalbazo;
\emph{On the nature of the so called generic instabilities in dissipative
relativistic hydrodynamics}, Gen. Relativ. Gravit. \textbf{41}, 1645-1654
(2009).

\bibitem[5]{Physica A 2009} A. Sandoval-Villalbazo, A.L. Garcia-Perciante,
L.S. Garcia-Colin; \emph{Relativistic transport theory for simple
fluids to first order in the gradients}, Phys. A. \textbf{388}, 3765-3770
(2009).

\bibitem[6]{ONSG}\foreignlanguage{english}{L. Onsager; Reciprocal
relations in irreversible processes I, Phys. Rev. 37, 405 (1931);
ibid 38 2265 (1931).}

\selectlanguage{english}%
\bibitem[7]{israel63}W. Israel; Relativistic Kinetic Theory of a
Simple Gas, J. Math. Phys. 4, 1163, (1963) .

\selectlanguage{spanish}%
\bibitem[8]{Garcia-Perciante Mendez Heat conduction ... revisited}
A. L. Garcia-Perciante, A. R. Mendez, \emph{Heat conduction in relativistic
neutral gases revisited}, Gen. Rel. Grav. \textbf{43}, 2257-2275,
(2011).

\bibitem[9]{ck}\foreignlanguage{english}{C. Cercignani and G. M.
Kremer, The relativistic Boltzmann equation: theory and applications,
Birkhauser, Basel, (2002).}

\bibitem[10]{benedicks}A.L. Garcia-Perciante, A. Sandoval-VIllalbazo,
L.S. Garcia-Colin; \foreignlanguage{english}{Benedicks effect in a
relativistic simple fluid, Jour. Non-Equilib. Thermodyn.} \textbf{38,}
141\textendash{}151 (2013).

\bibitem[11]{benoriginal}Benedicks, C., Ein fur Thermoelektrizitet
und metallische Warmeleitung fundamentaler Effekt, Ann. Phys., 360
(1918), 1\textendash{}80.

\selectlanguage{english}%
\bibitem[12]{degroor}de Groot, S. R., van Leeuwen, W. A. \& van der
Wert, Ch.; Relativistic Kinetic Theory, North Holland Publ. Co., Amsterdam
(1980).

\bibitem[13]{key-4}S. Weinberg; \textit{Gravitation and Cosmology},
Wiley (1972).

\selectlanguage{spanish}%
\bibitem[14]{JUTTNER} F. Juttner; Das Maxwellsche Gesetz der Geschwindigkeitsverteilung
in der Relativtheorie, Ann. Physik und Chemie, 34, 856, (1911).

\bibitem[15]{nosJNET-1}\foreignlanguage{english}{Garcia-Perciante
A. L., Sandoval-Villalbazo A., Garcia-Colin L. S, On the microscopic
nature of dissipative effects in special relativistic kinetic theory,
Jour. Non-Equilib. Thermodyn. \textbf{37} (2012) 43\textendash{}61.}

\selectlanguage{english}%
\bibitem[16]{remarks}A. L. Garcia-Perciante, A. Sandoval-Villalbazo;
Remarks on relativistic kinetic theory to first order in the gradients,
Journal of Non-Newtonian Fluid Mechanics 165:1024-1028, 2010.

\bibitem[17]{kremerschw}G. M. Kremer; Relativistic gas in a Schwarzschild
metric, J. Stat. Mech. 2013, 04016 (2013).

\bibitem[18]{tolman}A. Sandoval-Villalbazo, A. L. Garcia-Perciante,
D.Brun-Battistini; Tolman's law in linear irreversible thermodynamics:
a kinetic theory approach, Phys. Rev. D 86, 084015 (2012).\end{thebibliography}
\end{document}